\documentclass[aps,prl,twocolumn,superscriptaddress,showpacs]{revtex4}

\usepackage[dvips]{graphicx}
\usepackage{amsmath}
\usepackage{amssymb}
\usepackage{bm}

\begin{document}

\title{The Thermodynamic Glass Transition in Finite Dimensions}

\author{M.\ A.\ Moore}
\affiliation{School of Physics and Astronomy, University of Manchester,
Manchester M13 9PL, U.\ K.}
\author{J.\ Yeo}
\affiliation{Department of Physics, Konkuk University, Seoul 143-701, Korea}
\date{\today}

\begin{abstract}
Using an effective potential method, a replica formulism is set up for
describing supercooled liquids near their glass transition. The resulting
potential is equivalent to that for an Ising spin glass in
 a magnetic field. Results taken from the droplet picture of spin glasses are
then used to provide an explanation of the main features of fragile glasses.   
\end{abstract}



\maketitle

Structural glasses  have been studied extensively  at mean-field level
by  relating   them  to   infinite-range  p-spin  spin   glass  models
\cite{KTW}.   The dynamical equations  of these  spin glass  models are
similar  to those of  the mode-coupling  theory of  liquids \cite{MCT}.
These equations predict a  dynamical transition at a temperature $T_d$
below  which an  ergodicity  breaking occurs.   In the  infinite-range
p-spin glass models, a discontinuous thermodynamic 
 transition also takes place at a
temperature $T_K$  and is usually  identified with the  Kauzmann glass
transition \cite{Kauzmann}. $T_K<T_d$.  It is  expected that outside
the mean-field limit, there will be no genuine dynamical transition at
$T_d$ due to the fact that the metastable states which trap the system
dynamically are  unstable in  finite dimensions because  of activation
processes  over the  finite  free-energy barriers  which separate  the
metastable  states.    However,  it  is  commonly   assumed  that  the
thermodynamic glass transition $T_K$ could still exist in principle in
finite dimensional systems. This paper is essentially an investigation
of this question.

We  start   by  constructing  an  effective   Hamiltonian  $H[q]$  for
supercooled liquids  near the putative  thermodynamic glass transition
in  finite dimensions.   We  find that  $H[q]$  is given  by the  same
generic  functional  which  describes  the one-step  replica  symmetry
breaking transition  \cite{Ferrero} for  the p-spin spin  glass model.
Our derivation is based on the effective potential method developed by
Franz and Parisi \cite{FP} for spin glass models.  A similar effective
Hamiltonian to  ours was  used in Refs.~\cite{Dzero}  and \cite{Franz}
for structural glasses, but not explicitly derived. 

The next step  is the elucidation of the  properties of $H[q]$.  There
are  two coupling  constants in  the theory,  $w_1$ and  $w_2$  and at
mean-field level, the thermodynamic transition is discontinuous (first order)
 if their
ratio $w_2/w_1$ is greater than one and continuous when less than one.
There are arguments for a growing lengthscale in structural glasses as
the   temperature   is   reduced   (\cite{KTW},   \cite{Franz},   Ref.
\cite{Dzero}   and  references  therein,   and  Bouchaud and Biroli
\cite{BB}) and  experimental evidence for a lengthscale of several
interparticle diameters already exists 
\cite{berthier}. We feel  that it is more natural  to study parameters
appropriate to the continuous  transition rather than the first-order
transition in order to explain  a growing lengthscale. Most studies of
p-spin  models   have  focussed  on  parameters   which  generate  the
first-order  transition  to  the  one-step replica  symmetry  breaking
state, simply because  in that situation one also  has a mechanism for
explaining the  dynamical transition $T_d$.  However, it is  clear from
our derivation of the  p-spin functional that the $q_{\alpha\beta}$ in
it is  not simply  related to density  fluctuations so  our functional
cannot be used to directly  study  dynamical effects in density
correlations  functions.  In  our view,  they are  best  determined by
mode-coupling theory \cite{MCT}.  Our p-spin functional is appropriate
only for the study of the putative glass transition at $T_K$.

 Numerical studies  have been performed  on a p-spin model  with $p=3$
by  Parisi et al.  \cite{PPR}, which at mean-field level
  has a continuous transition \cite{MY}, (because for 
it the ratio $w_2/w_1<1$).  They found a
rapidly growing  lengthscale in this  p-spin model as  the temperature
was  reduced and  an  attempt was  even  made to  obtain its  critical
exponents in four dimensions.  Moore and Drossel \cite{MD} later showed that
the continuous transition,  if any, would be in  the same universality
class as  the Ising spin  glass in a  field -- the  transition usually
referred  to as  the  de Almeida-Thouless  (AT) transition  \cite{AT}.
However, it has been argued that the AT transition does not actually exists for
systems whose dimensionality is  less than six \cite{mam}, but instead
there is a crossover line in the field-temperature phase diagram where
the  growing correlations between  the spins  saturate.  The  value of
this  lengthscale  can be  estimated  from  droplet scaling  arguments
\cite{droplet}.  In  the numerical study  of Parisi et  al. \cite{PPR}
the largest  system size studied  was of linear dimension  six lattice
spacings and if the correlation length  is larger than the size of the
system then the  crossover would look like a  phase transition. In the
work  of   Moore  and  Drossel   \cite{MD}  the  lengthscale   in  the
(three-dimensional) Migdal-Kadanoff  approximation saturated at around
28 lattice  spacings at  low temperatures.  We  shall also  show using
droplet ideas borrowed from  spin glasses how this growing lengthscale
can produce dynamical behavior of the Vogel-Fulcher type \cite{VK}.

The   calculations   start   with   defining  an   overlap   $p_c({\bf
r})=\delta\rho_1({\bf    r})\delta\rho_2({\bf    r})$   between    two
configurations            of            density           fluctuations
$\delta\rho=\rho-\langle\rho\rangle$ in two  copies of the liquid.  By
averaging  over the  density configurations  of the  first  copy while
holding the density  configurations in the second copy  fixed at fixed
values of the overlap $p_c({\bf r})$ we obtain a constrained partition
function
\begin{equation}
Z[p_c({\bf      r}),\delta\rho_2({\bf      r})]=\langle\delta(p_c({\bf
r})-\delta\rho_1({\bf r}) \delta\rho_2({\bf r}))\rangle_{\rho_1}.
\label{partition}
\end{equation}
The effective  potential is  given by the  average of the  free energy
with  respect  to  the  density  configurations in  the  second  copy:
$\Omega[p_c({\bf                                      r})]=-T\langle\ln
Z[p_c,\delta\rho_2]\rangle_{\rho_2}$.  In order to perform the average
of  the  logarithm,   we  use  the  replica  method   and  write  $\ln
Z=\lim_{n\to 0}(Z^n-1)/n$.   Using the integral  representation of the
delta  function  in  (\ref{partition}),  we can  write  the  effective
potential as the following  averages over $\rho_2$ and $\rho_1^\alpha$
$(\alpha=1,\cdots,n)$:
\begin{eqnarray}
&&\Omega[p_c({\bf
r})]=-T\int\prod_{\alpha}\frac{\mathcal{D}\lambda_\alpha}{2\pi}\;
\exp\left[  i\sum_{\alpha}\int  d{\bf   r}  \lambda_\alpha  ({\bf  r})
p_c({\bf  r})   \right]  \nonumber  \\   &&\quad  \times  \left\langle
\left\langle        \exp\left[-i\sum_\alpha\int        d{\bf        r}
\delta\rho_1^\alpha({\bf  r})\delta\rho_2({\bf  r})\lambda_\alpha({\bf
r})\right] \right\rangle_{\rho_2}\right\rangle_{\rho_1^\alpha}.
\end{eqnarray}
 
We evaluate  the above averages  over $\rho_1^\alpha$ and  $\rho_2$ by
making successive  cumulant expansions. The resulting  integral can be
expressed in  terms of the liquid-state  correlation functions $G({\bf
r}_1,{\bf       r}_2)=\langle\delta\rho({\bf      r}_1)\delta\rho({\bf
r}_2)\rangle$,           $G({\bf          r}_1,{\bf          r}_2,{\bf
r}_3)=\langle\delta\rho({\bf         r}_1)\delta\rho({\bf        r}_2)
\delta\rho({\bf   r}_3)\rangle$,    etc.    Keeping   terms    up   to
$O(\lambda^3)$, we have
\begin{eqnarray}
&&\Omega[p_c({\bf                                            r})]\simeq
-T\int\prod_{\alpha}\frac{\mathcal{D}\lambda_\alpha}{2\pi}\;
\exp\left[  i\sum_{\alpha}\int  d{\bf   r}  \lambda_\alpha  ({\bf  r})
  p_c({\bf r})  \right] \nonumber \\  &&\times\exp\left[-\frac 1 2\int
  d1 d2  \; G^2(1,2) \sum_\alpha  \lambda_\alpha (1)\lambda_\alpha (2)
  \right] \nonumber \\  &&\times\exp\left[\frac i 6\int  d1 d2 d3  \; G^2(1,2,3)
  \sum_\alpha   \lambda_\alpha   (1)\lambda_\alpha   (2)\lambda_\alpha
  (3)\right] \nonumber\\ &&\times \exp\left[\rm{O}(\lambda_{\alpha}^4,
  \lambda_{\alpha}^2\lambda_{\beta}^2)\right],
\end{eqnarray}
where  $1,2,\cdots$  denote   the  position  vectors  ${\bf  r}_1,{\bf
r}_2,\cdots$.

Now    we    define   $q_{\alpha\beta}({\bf    r})=\lambda_\alpha({\bf
r})\lambda_\beta({\bf  r})$   for  $\alpha\neq\beta$  and   write  the
effective   potential  as   a  functional   integral  of   the  fields
$q_{\alpha\beta}$.  We insert the  following identity inside the above
expression:
\begin{eqnarray}
&&1=                 \int\prod_{\alpha<\beta}\mathcal{D}q_{\alpha\beta}
\int\prod_{\alpha<\beta}\frac{\mathcal{D} u_{\alpha\beta} }{2\pi}\; \\
&&\times       \exp\left[i\sum_{\alpha<\beta}\int      d{\bf      r}\;
u_{\alpha\beta}({\bf           r})          \left(q_{\alpha\beta}({\bf
r})-\lambda_\alpha({\bf     r})\lambda_\beta({\bf    r})\right)\right].
\nonumber
\end{eqnarray}
Due to the  coupling between $\lambda_{\alpha}$ and $u_{\alpha\beta}$,
the  integration over  $\lambda_{\alpha}$ generates  various  terms in
$u_{\alpha\beta}$.  For  example, to  the lowest order,  the quadratic
term in $u_{\alpha\beta}$ comes  from joining two pairs of $\lambda$'s
in the  two $u\lambda\lambda$  vertices. Therefore the  coefficient of
the quadratic term in $u_{\alpha\beta}$ involves, to the lowest order,
the square of the $\lambda$-propagator $K(1,2)$ satisfying
\begin{equation}
\int d3 K(1,3)\left[ G^2(3,2) \right]=\delta(12).
\end{equation}
There are, of course, other  contributions to the $O(u^2)$ term, which
come  from  the  higher  order  vertices  in  $\lambda$.   These  just
renormalize   its  coefficient.    Two   kinds  of   cubic  terms   in
$u_{\alpha\beta}$ are generated, namely, $\mathrm{Tr}\; u^3$ and $\sum
u^3_{\alpha\beta}$.  The  coefficients of these terms  will contain in
general the contributions from the higher order vertices in $\lambda$.
Keeping up to the cubic order in $u_{\alpha\beta}$, we have
\begin{eqnarray}
&&\Omega[p_c({\bf                                              r})]\sim
-T\int\prod_{\alpha<\beta}\mathcal{D}q_{\alpha\beta}
\int\prod_{\alpha<\beta}\frac{\mathcal{D}   u_{\alpha\beta}  }{2\pi} \\
&&\times\exp\big[i\sum_{\alpha<\beta}\int      d{\bf      r}\;u_{\alpha\beta}({\bf
    r})q_{\alpha\beta}({\bf r})\big]  \nonumber \\ 
&&\times\exp\big[\frac  i 2
  \sum_{\alpha<\beta}\int
  d1d2d3\;A(1,2,3)u_{\alpha\beta}(1)p_c(2)p_c(3)  \big]   \nonumber\\
&&\times\exp\big[-\frac     1    2
  \sum_{\alpha<\beta}\int                d1d2\;B(1,2)u_{\alpha\beta}(1)
  u_{\alpha\beta}(2)\big]    \nonumber   \\    
&&\times\exp\big[\frac    i   6
  \sum_{(\alpha,\beta,\gamma)}\int                     d1d2d3\;C(1,2,3)
  u_{\alpha\beta}(1)u_{\beta\gamma}(2)u_{\gamma\alpha}(3)  \big]\nonumber\\
&&\times\exp\big[\frac  i 6
  \sum_{\alpha<\beta}\int            d1d2d3\;D(1,2,3)u_{\alpha\beta}(1)
  u_{\alpha\beta}(2)u_{\alpha\beta}(3)\big], \nonumber
\end{eqnarray}
where  we  dropped  the  terms   in  $p_c$  which  do  not  couple  to
$u_{\alpha\beta}$.  The coefficients $A,B,C$ and $D$ are given, to the
lowest order, by
\begin{eqnarray}
&&A(1,2,3)\simeq K(1,3)K(2,3)  , \label{A} \\  
&&B(1,2)\simeq K^2(1,2) ,\label{B}\\   
&&C(1,2,3)\simeq   K(1,2)K(2,3)K(3,1)   ,\label{C}   \\
&&D(1,2,3)\simeq -\int\prod_{i=4}^9 d{\bf r}_i\; G^2(4,5,6)G^2(7,8,9)
\label{D} \\ 
&&\quad \times K(1,4)K(1,7)K(2,5)K(2,8)K(3,6)K(3,9).
\nonumber
\end{eqnarray}

Finally we  integrate over  $u_{\alpha\beta}$ to obtain  the effective
Hamiltonian in terms of  $q_{\alpha\beta}$ as follows. First note that
in addition  to the usual  quadratic term in  $q_{\alpha\beta}$, there
are  cubic terms  in $q$  in the  form $\mathrm{Tr}\,  q^3$  and $\sum
q^3_{\alpha\beta}$  with real  coefficients.   We also  note that  the
terms  involving  $p_c$  couple   to  $q_{\alpha\beta}$  in  the  form
$p_c^{2k}q_{\alpha\beta}^{k^\prime}$  with integers $k$  and $k^\prime
$.    The   resulting  expression   gives   the  effective   potential
$\Omega[p_c]$  in  terms of  a  functional  integral  over the  fields
$q_{\alpha\beta}$. Since  the value of the overlap  is determined from
the condition $\delta\Omega/\delta  p_c=0$, $p_c({\bf r})=0$ is always
a solution and describes the  liquid phase.  We take the effective
Hamiltonian     $H[q]$   for    supercooled     liquids    as
$\Omega[p_c=0]\sim\int\prod_{\alpha<\beta}\mathcal{D}q_{\alpha\beta}\exp[-H[q]]$.
The coefficients  of various terms in $q_{\alpha\beta}$  in $H[q]$ are
in    general    expressed   in    a    nonlocal    way   much    like
Eqs. (\ref{A})-(\ref{D}).   However, the  most relevant terms  will be
the local ones where all the  points in space are the same. Keeping as
usual   the  derivative   term  only   in  the   quadratic   order  in
$q_{\alpha\beta}$, we obtain the effective Hamiltonian as
\begin{eqnarray}
H[q]&=&   \int  d{\bf   r}\;\Big\{  \frac   c   2  \sum_{\alpha<\beta}
\left(\nabla     q_{\alpha\beta}({\bf     r})\right)^2    +\frac{t}{2}
\sum_{\alpha<\beta}    q^2_{\alpha\beta}({\bf    r})   \nonumber    \\
&&-\frac{w_1}{6}    \mathrm{Tr}    \;q^3({\bf    r})    -\frac{w_2}{3}
\sum_{\alpha<\beta} q^3_{\alpha\beta}({\bf r})\Big\}.
\label{HF}
\end{eqnarray}
The  coefficients  $c,t ,w_1$  and  $w_2$  will  be functions  of  the
temperature and density of the liquid, with smooth dependence on them.
 The study  of the first-order transition which
occurs at mean-field level requires  the inclusion of terms quartic in
the $q_{\alpha\beta}$ fields, but  these are \lq irrelevant' variables
at the  presumed continuous transition  in finite dimensions  and they
will be dropped. At mean-field  level, the transition is continuous if
$w_2/w_1 \leq  1$ \cite{Ferrero}. In the  p-spin replicated functional
of Eq. (\ref{HF})  the coefficients in it have  contributions from all
the higher order  cumulants of the density etc..  Fortunately as we are
studying  a continuous  transition, their  numerical values  will only
effect non-universal features of the putative transition.

The thermal average $\langle q_{\alpha\beta} \rangle$ is non-zero even
above  any supposed transition  (when one  goes beyond  the mean-field
approximation)  because of  the term  in $w_2$  in the  Hamiltonian of
Eq. (\ref{HF}).   As a consequence the only  possible continuous phase
transition  is   one  which  breaks   the  replica  symmetry   of  the
high-temperature phase.   This transition was  shown in Ref.~\cite{MD}
to be in  the same universality class as that  for the transition from
the paramagnet to the  spin glass phase in a field,  first discussed by de
Almeida and  Thouless \cite{AT}.  The Hamiltonian of an Ising spin
glass in a field is
\begin{equation}
\mathcal{H}=-\sum _{<ij>} J_{ij}S_iS_j -h\sum _i S_i.
\label{SGH}
\end{equation}
The  couplings  $J_{ij}$  are  quenched  random  variables  connecting
(usually)   nearest-neighbor   sites.    The  Edwards-Anderson   order
parameter is  $\tilde{q}= \sum_i\langle  S_i \rangle^2/N$ and  in zero
field becomes  non-zero below $T_s$,  the critical temperature  of the
zero-field spin glass transition. Using the replica method one can set
up  a functional  analogous  to  that for  structural  glasses in  Eq.
(\ref{HF}) for the spin  glass Hamiltonian \cite{BR}.  It differs only
in that the cubic term  $(w_2/3)\sum \, q^3_{\alpha\beta}$ is absent but
there is instead a linear  term $h^2\sum \, q_{\alpha\beta}$. Now when
the temperature $T$ is in the vicinity of the thermodynamic structural
\lq glass transition' $T_K$,  we expect it to be well  below $T_s$.  At such
temperatures  $w_2$ and  $h^2$ can be related by
noting  that the  cubic  term  could be  approximated  by $w_2\sum  \,
\langle   q^2_{\alpha\beta}\rangle    q_{\alpha\beta}   $.    $\langle
q^2_{\alpha\beta}\rangle$ will be independent of $\alpha $ and $\beta$
as  we are working  in the  replica symmetric  phase, and  only slowly
varying as a  function of temperature so then  $h^2\propto w_2$.  Then
because  $w_2$ is a  function of  the temperature  and density  of the
liquid,  $h^2$  in  the Ising  spin  glass  analogue  will also  be  a
temperature and density-dependent quantity.

The question  then of whether  there is or  is not a  structural glass
transition then turns on whether there is an AT line in the spin glass
analogue.   This has  been controversial  in the  past,  but numerical
\cite{YK} and experimental data  \cite{ND} favors its absence in three
dimensional spin glasses and we further argued in Ref.~\cite{mam} that
for  all  dimensions  $d<6$ there  would  be  no  AT line.   We  shall
henceforth take it that there is no AT line and hence that there is no
thermodynamic  glass transition  for three  dimensional  systems.  How
then can  we explain  the phenomena normally  associated with  the
glass transition?

The key to  understanding this point is to realize  that even if there
is no AT line there can be a growing lengthscale as the temperature is
reduced.  The  lengthscale is analogous  to the \lq mosaic'  lengthscale of
other authors \cite{KTW, Franz, Dzero, BB}, but its physical origin is
different.  In  our case,  the lengthscale is
equal to the  size of the droplet in the spin  glass analogue which is
typically  flipped by the  application of  the field  $h$.  To  flip a
region  of size  $L$  in the  low-temperature  phase of  a spin  glass
typically   costs   an   \lq   interface'   free   energy   of   order
$(L/\xi(T))^{\theta}$,  where  $\xi(T)$   is  the  correlation  length
associated with  the spin glass transition in  zero-field and $\theta$
is an exponent typically  $\sim 0.2$ \cite{droplet}.  Each droplet has
a magnetic moment of order  $\sqrt{\tilde{q}}L^{d/2}$ so in a field of
magnitude $h$ the domains are of size $L^{*}$, where
\begin{equation}
(L^{*}/\xi(T))^{\theta} \sim \sqrt{\tilde{q}}(L^{*})^{d/2}h,
\label{SM}
\end{equation}
in order that the energy gained by flipping a droplet in the field $h$
just balances the interface free energy cost.  This result for $L^{*}$
should  be valid provided  $L^{*}>\xi(T)$.

In  the  droplet picture  the  free-energy  barrier  which has  to  be
surmounted   in  order   to  flip   a  region   of  size   $L^{*}$  is
$A(L^{*}/\xi(T))^\psi$,  where  $A$  is   of  order  of  the  critical
temperature $T_s$ \cite{droplet}.  The basic relaxation time $\tau$ in
the system is of Arrhenius form:
\begin{equation}
\tau \sim\tau_0 \exp[A(L^*/\xi(T))^{\psi}/T].
\label{tau}
\end{equation}
 Alas, the value  of $\psi$ in three dimensional  spin glasses is very
 poorly determined, with quoted values varying widely.

When  $T \ll  T_s$,  $\xi(T)$ will  be  a slowly  varying function  of
temperature  and of  order of  the interparticle  spacing. $\tilde{q}$
will also  then be nearly  temperature-independent.  If $w_2$  is only
weakly   temperature   dependent,   $L^{*}$  will   be   approximately
temperature  independent  and  $\tau$  will  then  be  of  the  simple
Arrhenius form.   This will be the  case then of  strong glasses.  For
fragile glasses we expect that
\begin{equation}
w_2 \sim h^2 \sim (T-T_0),
\label{Taylor}
\end{equation}
  so that 
$w_2$ appears to be going to zero at $T_0$. We stress that
 the effective value of
$w_2$ probably will never vanish, since even if the bare
 coefficient did go to zero the higher terms in the 
 functional $H[q]$ which break the time-reversal invariance
 will make its effective value non-zero so Eq. (\ref{Taylor}) 
is just a simple phenomenological way to obtain a growing lengthscale.    Then
\begin{equation}
\tau                            \sim                            \tau_0
\exp\Big[\frac{C}{(T-T_0)^{\frac{\psi}{d-2\theta}}}\Big],
\end{equation} 
where $C$ denotes  a constant. It is clear  that this has similarities
to the Vogel-Fulcher \cite{VK} form for structural relaxation times in
the supercooled liquid:
\begin{equation}
\tau \sim \tau_0 \exp\Big[\frac{C}{T-T_0}\Big].
\end{equation} 
 Note that we have proceeded on the assumption that the exponent
$\psi$ for the supercooled liquid and  the spin glass are the same. In
principle our  mapping between  liquids and spin glasses
 is explicit  only for
\lq  equilibrium'  quantities  and   might  not  extend  to  dynamical
features. Note that the droplet formula for $\tau$ of Eq.~ (\ref{tau}) is only
expected to hold when $L^{*}$ is much larger than the typical distance between
the molecules of the glass,
 and experiments might not have reached the region where it
has become valid; in other words the simple exponent of the Vogel-Fulcher fit
might reflect  pre-asymptotic behavior.

The simulations of Parisi et  al. \cite{PPR} show that the entropy per
spin  went  to zero apparently  as  $s_c(T)  \approx  k_B (T-T_K)$, although 
no data was obtained close to $T_K$ and without 
a genuine transition the entropy is unlikely
to ever vanish completely.  A  similar
behavior  is found  for the  configurational entropy  per  particle in
fragile  glasses. What  is striking  about  the simulations  is that  the
temperature $T_K$  was very  close to the  temperature $T_0$  at which
the correlation  length was apparently diverging  (the difference was
less than  1\%).  One  feature common to  fragile glasses is  that for
most of  them $T_0$ extracted from  the Vogel-Fulcher fit  is close to
the  Kauzmann   temperature  $T_K$,  the  temperature   at  which  the
extrapolated  entropy of  the supercooled  liquid apparently  falls to
that  of the  crystalline  phase.  The  ratio  $T_K/T_0$ lies  between
0.9--1.1 for many  glass formers amongst which $T_K$  ranges from 50 K
to  1000  K  \cite{Angell}. Since we 
are arguing that there is no genuine transition
but only a crossover, the
\lq$T_0$' obtained from the timescale $\tau$,
the $T_0$ of Eq. (\ref{Taylor}) and $T_K$ need not be precisely the same.   

We should like  to acknowledge the hospitality of  the Korea Institute
for  Advanced Study  during  the  preparation of  this  paper. We  are
indebted  to Drs G.  Biroli and  J-P Bouchaud  for their  comments and
insights.

\end{document}